\documentclass{amsart}
\usepackage{amssymb}
\usepackage{amsmath}
\usepackage{amsfonts}
\usepackage{graphicx}
\usepackage{epstopdf}

\usepackage{ytableau}
\usepackage[vcentermath]{youngtab}

\usepackage{float}
\usepackage{hyperref}

\newcommand{\beq}{\begin{equation}}
	\newcommand{\eeq}{\end{equation}}

\begin{document}
	
	\begin{center}
		{\Large\bf Universal Quantum Dimensions: I. 
			$\gamma$-Independent Factors }\\
		\vspace*{1 cm}
		{\large  }
		\vspace*{0.5 cm}
		
		{\large  R.L.Mkrtchyan 
		}

		\vspace*{0.5 cm}
		
		{\small\it Alikhanyan National Science Laboratory (Yerevan Physics Institute), \\ 2 Alikhanian Br. Str., 0036 Yerevan, Armenia}
		
		{\small\it E-mail: mrl55@list.ru}
		
		\vspace*{0.5 cm}

	\end{center}

	{\bf Abstract.}  We propose a method for computing universal (in Vogel's sense) quantum dimension formulae for universal multiplets whose associated $sl$, $so$, and $sp$ representations are nonzero. The method uses the relation between $sl$ and $so$ representations given by the vertical-sum operation, and the dual relation between $sl$ and $sp$ representations given by the horizontal-sum operation on the corresponding Young diagrams. The usual quantum dimensions of these three representations, together with subtleties related to the invariance of universal formulae under automorphisms of the $sl$ Dynkin diagram, allow one to determine the $\gamma$-independent factors of a universal quantum dimension (note that  $\gamma$ is the only parameter for classical algebras, depending on their rank). Using this approach, we compute the $\gamma$-independent factors for (known) adjoints' universal quantum dimension, and also obtain such factor in one new case. We discuss how to extend this approach to the  $\gamma$-dependent factors in the quantum dimension formulae, and other issues. This is another 	instance in which calculations purely within the classical algebras predict the 	answers for the exceptional cases, due to the hidden universality structure of 	the theory of simple Lie algebras.

	\vspace*{1 cm}

	\section{Introduction}

	The aim of the present paper is to develop a new method for computing universal (quantum) dimension formulae. It builds on the method for computing universal Casimir eigenvalues suggested in our previous paper \cite{Mkrtchyan2025b}. Other universal quantities, such as Chern--Simons partition functions or the universal volume of groups, lie outside the scope of the present work. 
	
	The quantum dimension of a representation of a simple Lie algebra with highest weight $\lambda$ is usually defined as the value of its character on the Weyl line $x\rho$, where $x$ is a parameter and $\rho$ is the Weyl vector (the sum of fundamental weights, equivalently half the sum of all positive roots). The ordinary dimension is recovered from the quantum dimension in the limit $x \rightarrow 0$. The Weyl character formula leads to the following expression for the quantum dimension (see e.g. \cite{DiFrancesco1997}, eq. (13.170)):
	
	\begin{eqnarray}\label{qWeyl}
		dim_q(\lambda)=\chi_\lambda (x\rho)= \prod_{\alpha > 0} \frac{\sinh((\alpha,\lambda+\rho)x/2)}{\sinh((\alpha,\rho)x/2)}
	\end{eqnarray}
	
	It turns out that for many representations this formula can be written as a product/ratio of hyperbolic sines whose arguments are linear functions of the universal parameters. For example, for the quantum dimension of the adjoint representation one has \cite{Westbury2003,MkrtchyanVeselov2012}:

	\begin{eqnarray}  \label{cad}
		dim_q(\mathfrak{g}) &=& -\frac{\sinh\left(\frac{(\gamma+2\beta+2\alpha)x}{4}\right)}{\sinh\left(\frac{\gamma x}{4}\right)}\frac{\sinh\left(\frac{(2\gamma+\beta+2\alpha)x}{4}\right)}{\sinh\left(\frac{\beta x}{4}\right)}\frac{\sinh\left(\frac{(2\gamma+2\beta+\alpha)x}{4}\right)}{\sinh\left(\frac{\alpha x}{4}\right)}
	\end{eqnarray}

	Another example is the quantum dimension of the square of the adjoint. It decomposes into symmetric and antisymmetric parts, and both admit universal expressions. For the symmetric part one has the universal decomposition \cite{Vogel1999}
	
	\begin{eqnarray}\label{sad}
		S^2 \mathfrak{g}=1 \oplus Y_2(\alpha) \oplus Y_2(\beta) \oplus Y_2(\gamma)
	\end{eqnarray}
	
	with quantum dimensions of the summands given by
	
	\begin{eqnarray} \label{Y2c}
		& dim_q(Y_2(\alpha)) =  \\ \nonumber
		&  -\frac{\sinh[\frac{x t}{2}]\sinh[\frac{x(\beta-2t)}{4}]\sinh[\frac{x(\gamma-2t)}{4}]\sinh[\frac{x(\beta+t)}{4}]\sinh[\frac{x(\gamma+t)}{4}]\sinh[\frac{x(3\alpha-2t)}{4}]}{\sinh[\frac{x\alpha}{4}]\sinh[\frac{x\alpha}{2}]\sinh[\frac{x\beta}{4}]\sinh[\frac{x\gamma}{4}]\sinh[\frac{x(\alpha-\beta)}{4}]\sinh[\frac{x(\alpha-\gamma)}{4}]}
	\end{eqnarray}
	and similarly for $Y_2(\beta)$ and $Y_2(\gamma)$ by permutation of parameters. 
	
	Likewise, the antisymmetric square decomposes universally as
	
	\begin{eqnarray}
		\label{decompchar2}
		\wedge^2(\mathfrak{g})= \mathfrak{g}+X_2
	\end{eqnarray}
	where $X_2$ is an irreducible representation for all algebras, provided we extend them by automorphisms of the corresponding Dynkin diagrams---in particular, by $\mathbb{Z}_2$ in the case of $sl$ algebras. 
	
	Quantum dimension of representation $X_2$ is \cite{Mkrtchyan2017}
	
	\begin{eqnarray}\label{qX2}
		&
		dim_q(X_2)=\frac{\sinh\left[\frac{(2t-\alpha)x}{4}\right]\sinh\left[\frac{(2t-\beta)x}{4}\right]\sinh\left[\frac{(2t-\gamma)x}{4}\right]}{\sinh\left[\frac{\alpha x}{4}\right]\sinh\left[\frac{\beta x}{4}\right]\sinh\left[\frac{\gamma x}{4}\right]}  \times \\ \nonumber
		&
		\frac{\sinh\left[\frac{(t+\alpha)x}{4}\right]\sinh\left[\frac{(t+\beta)x}{4}\right]\sinh\left[\frac{(t+\gamma)x}{4}\right]}{\sinh\left[\frac{\alpha x}{2}\right]\sinh\left[\frac{\beta x}{2}\right]\sinh\left[\frac{\gamma x}{2}\right]}   
		\frac{\sinh\left[\frac{(t-\alpha)x}{2}\right]\sinh\left[\frac{(t-\beta)x}{2}\right]\sinh\left[\frac{(t-\gamma)x}{2}\right]}{\sinh\left[\frac{(t-\alpha)x}{4}\right]\sinh\left[\frac{(t-\beta)x}{4}\right]\sinh\left[\frac{(t-\gamma)x}{4}\right]}
	\end{eqnarray}

	Universal multiplets are defined in \cite{Mkrtchyan2025b}. For a given universal dimension formula, we call the {\it small universal multiplet for a given simple Lie algebra} the set of representations obtained by evaluating that formula at the values of universal parameters for that algebra in Vogel's Table \ref{tab:Vogel}, together with their simultaneous permutations. Thus a small multiplet has at most six members, although most of them are usually zero. The {\it big universal multiplet of a given universal dimension formula} is the union of the small multiplets over all simple Lie algebras.

	With this terminology, for the dimension formula (\ref{cad}) the small multiplets consist of the adjoint representation (permutations give the same representation), and the big multiplet is the union of all adjoint representations. The square of the adjoint representation contains several universal multiplets. Its antisymmetric part contains, first, by (\ref{decompchar2}) the universal big multiplet of adjoint representations. The other multiplet is the big multiplet of $X_2$ representations; for each algebra it is the only nonzero member of the corresponding small multiplet, similarly to the adjoint case. The symmetric square of the adjoint consists of two big multiplets. One is the singlet, i.e. its small multiplets consist of a single singlet for each algebra. The other is the big multiplet of $Y_2(\cdot)$ representations; the small multiplets for classical groups are listed in Table \ref{tab:ad2}. 
	
	\ytableausetup{smalltableaux}
	
	\begin{table}
		\caption{Universal decomposition of the square of adjoint}
		\label{tab:ad2}
		
		\begin{tabular}{|c|c|c|c|c|}
			\hline
			Irrep	&  Casimir&$sl$  &$so$  & $sp$ \\ 
			\hline
			$\mathfrak{g}$	&  $2t$ &  $D_s\left( \ydiagram{1},\ydiagram{1} \right)$  &  \ydiagram{1,1} & \ydiagram{2} \\
			\hline
			$Y_2(\alpha)$	& $4t-2\alpha$ & $D_s\left(  \ydiagram{2},\ydiagram{2} \right) $   & \ydiagram{2,2} & \ydiagram{4} \\
			\hline
			$Y_2(\beta)$	& $4t-2\beta$ &$D_s \left(   \ydiagram{1,1},\ydiagram{1,1} \right) $ & \ydiagram{1,1,1,1}  & \ydiagram{2,2} \\
			\hline
			$Y_2(\gamma)$	& $4t-2\gamma$ & $ D_s \left( \ydiagram{1},\ydiagram{1}\right)$ &  \ydiagram{2}&  \ydiagram{1,1} \\
			\hline
			$X_2$	& $4t$ &$D_s\left( \ydiagram{2},\ydiagram{1,1} \right)$& \ydiagram{2,1,1} & \ydiagram{3,1}  \\
			\hline
		\end{tabular}
	\end{table}

	In Table \ref{tab:ad2} we use the notation $D(\lambda,\tau)$ for $sl(N)$ representations with Dynkin labels
	\begin{eqnarray}
		D(\lambda,\tau)=(\lambda_1, \lambda_2, ...,\lambda_k, 0,...,0, \tau_k,...,\tau_2,\tau_1)
	\end{eqnarray}
	with fixed (i.e., $N$-independent) $k$.  
	According to \cite{Deligne1996,CohenMan1996} one needs the representations symmetrized with respect to the $\mathbb{Z}_2$ automorphism of the Dynkin diagram:
	$D_s(\lambda,\tau)=D(\lambda,\tau)\oplus D(\tau,\lambda)$ for $\lambda \neq \tau$, and $D_s(\lambda,\lambda)=D(\lambda,\lambda)$.  
	We shall also use the notation $(\lambda_1,\lambda_2,\dots,\lambda_k,0,\dots,0,\tau_k,\dots,\tau_1)_s=D_s(\lambda,\tau)$.  
	The labels $(\lambda_1,\lambda_2,\dots)$  and $(\tau_1,\tau_2,\dots)$  correspond in a standard way to the Young diagrams $\lambda$ and $\tau$.  
	We also assume that $N$ is “large enough’’ so that the antisymmetric tensors involved do not vanish.

	According to \cite{Mkrtchyan2025b}, Vogel's Table \ref{tab:Vogel} induces, within each big multiplet, a correspondence between individual representations of different algebras, i.e. between specific members of the small multiplets. Namely, we call the representations obtained by evaluating a universal dimension formula at the values in Vogel's Table \ref{tab:Vogel} the {\it associated representations for that formula}, and likewise for representations arising from simultaneous, for all algebras, permutations of the universal parameters.  
	In Table \ref{tab:ad2}, the associated representations are those appearing in each row.

	How can one derive expressions for universal dimensions and universal quantum dimensions? Vogel originally derived universal formulae (in particular, the decomposition \ref{sad}) by manipulating Jacobi diagrams using the IHX and AS relations (speaking on the knot theory language). He defined the universal parameters $\alpha, \beta, \gamma$ in terms of Casimir eigenvalues of representations $Y_2(\cdot)$, which parameterize the universal Lie algebra \cite{Vogel2011}: in particular, any Jacobi diagram can be expressed in terms of these parameters. They parameterize simple Lie algebras via Vogel's Table \ref{tab:Vogel}. However, this method becomes unwieldy for higher powers of the adjoint, and quantum dimensions are outside the scope of that approach.
	
	Another approach to deriving universal formulae is based on root-system considerations \cite{LandsbergManivel2006,Mkrtchyan2017}. It also allows one to derive quantum dimensions. However, for higher powers of the adjoint, root data appear too limited to reconstruct the universal dimensions. 
	
	A different approach, developed in \cite{IsaevKrivonosProvorov2023}, is based on the simultaneous use of (conjectural) universal (split) Casimir eigenvalues and equations for the dimensions of representations coming from traces of powers of the Casimir operator. Universal quantum dimensions seem to be outside the scope of this method, and it becomes computationally difficult for higher powers of the adjoint.

	The goal of the present paper is to develop a new method for deriving universal (quantum) dimension formulae. At present, the method is restricted to universal multiplets with nonzero associated $sl$, $so$, and $sp$ representations. The derivation is based on the connection between $sl$ and $so$ small multiplets discovered in \cite{Mkrtchyan2025b}: the Young diagram of the $so$ member is given by the so-called vertical sum of the Young diagrams of the corresponding $sl$ representation. In the present paper we also use the dual version of this connection, namely the {\it horizontal sum} of Young diagrams. The corresponding statement concerns $sp$ algebras: in the universal multiplets there are associated $sl$ and $sp$ members such that the $sp$ representation has Young diagram given by the horizontal sum of the two Young diagrams defining the $sl$ member.

	Knowing the ordinary quantum dimensions of these three representations, one may attempt to reconstruct the corresponding universal quantum dimension. In the present paper, we demonstrate, on examples, how this can be done for the $\gamma$-independent part of the universal quantum dimension, namely the factor that does not involve the parameter $\gamma$ from Vogel’s Table~\ref{tab:Vogel}. Recall that $\gamma$ is the only Vogel parameter that depends on the rank of the algebra. Quantum dimensions play a crucial role here, as they allow for a clean separation of different contributions. We show, in particular, that it is essential to take into account the invariance of universal formulae under automorphisms of the $\mathfrak{sl}$ Dynkin diagram.
	
	Altogether, the proposed approach consists of the following steps. One first chooses an $\mathfrak{sl}$ representation $D_s(\lambda,\tau)$ satisfying certain restrictions; for instance, the Young diagrams $\lambda$ and $\tau$ should have equal areas, say both equal to $A$. This is a necessary condition for the representation to appear in the decomposition of a power $\mathfrak{g}^{\otimes A}$ of the adjoint representation. If such a representation occurs in $\mathfrak{g}^{\otimes A}$ “alone,” in the sense that there is no other representation with the same Casimir eigenvalue, then one may assume \cite{AvetisyanIsaevKrivonosMkrtchyan2024} the existence of a universal multiplet equipped with a universal (quantum) dimension formula. Assuming their existence, the associated $\mathfrak{so}$ and $\mathfrak{sp}$ members of this multiplet can then be obtained via vertical and horizontal sum operations, respectively. One can subsequently compute the quantum dimensions of these three representations and, guided by the examples discussed below, reconstruct the $\gamma$-independent factor, and, hopefully, also the $\gamma$-dependent part, thus obtaining the complete universal formula.
	
	Some of these steps involve certain assumptions and ambiguities, and therefore each of them requires further analysis and refinement. In what follows, we present explicit calculations that support the key step of this program: the possibility of reconstructing the universal quantum dimension formula from the quantum dimensions of the associated $\mathfrak{sl}$, $\mathfrak{so}$, and $\mathfrak{sp}$ representations, illustrated through several examples.

	\section{The general form of universal quantum dimensions}

	The general form of the quantum dimension of an irreducible representation of a simple Lie algebra, computed from (\ref{qWeyl}), is 
	
	\begin{eqnarray}\label{gform}
		\prod_{i=1}^{k} \frac{\sinh(xa_i)}{\sinh(xb_i)}
	\end{eqnarray}
	where $a_i, b_i$ are rational numbers depending on the representation and the algebra.  The universal quantum dimensions have the same general form, with $a_i, b_i$ now being the linear functions of universal parameters, as in the examples above.

	Universal formulae yield (quantum) dimensions of irreducible representations of the original simple Lie algebra extended by the action of automorphisms of its Dynkin diagram \cite{Deligne1996,CohenMan1996}. In particular, if an $sl$ representation is invariant under the $\mathbb{Z}_2$ automorphism, then universal formulae give its (quantum) dimension coinciding with (\ref{qWeyl}). If a representation is not invariant under this $\mathbb{Z}_2$, then universal formulae instead give the dimension of the sum of the representation and its image under the nontrivial automorphism. Since their quantum dimensions coincide, the universal formula acquires an additional factor $2$. However, such a coefficient cannot appear from the Weyl formula, so how the universal formulae maintain the same form (\ref{gform}) both for the $Z_2$ invariant representations, and for the sum of two $Z_2$ non-invariant representations?
	
	As an example, consider the universal quantum dimension of $X_2$ (\ref{qX2}). When specialized to $sl$, it is the sum of two $Z_2$ non-invariant representations, connected by the non-trivial automorphism. There is no explicit factor $2$ in (\ref{qX2}), and at first sight there seems to be no way to obtain it from a product of sines. The point is that one factor of this universal expression, namely 
	
	\begin{eqnarray}
		\frac{\sinh\left[\frac{(t-\gamma)x}{2}\right]}{  \sinh\left[\frac{(t-\gamma)x}{4}\right]} 
	\end{eqnarray}
	
	specializes for $sl$ (i.e. at $\alpha+\beta=0$) to $2$, since both sines are zero and one has to take a limit. For other algebras this factor gives a nontrivial ratio of sines compatible with the Weyl formula (\ref{qWeyl}). Below we therefore assume the general form of a universal quantum dimension as a product/ratio of hyperbolic sines, as in (\ref{qX2}), and as in all other known universal formulae. 
	
	\section{The vertical sum relation of sl and so small multiplets}

	We refer to \cite{Mkrtchyan2025b} for details on definitions and statements related to the vertical sum of Young diagrams. The statement is that in the universal multiplets some associated $sl$ and $so$ members of corresponding small multiplets are connected by the vertical sum operation. Namely, if the $sl$ representation is $D_s(\lambda,\tau)$, then the corresponding $so$ member has Young diagram $\lambda\oplus_v \tau$, defined as follows. 
	
	Place Young diagrams $\lambda$ and $\tau$ one under another, aligning their left vertical boundaries.  
	Then push each box of the lower diagram upward until it either meets a box of the upper diagram or reaches the upper horizontal boundary of the upper diagram.  
	The result is the Young diagram $\lambda \oplus_v \tau$, called the {\it vertical componentwise sum} of $\lambda$ and $\tau$.
	
	Equivalently, one can define the same operation as:
	
	1. Let $h^\mu_i$ be the height of the $i$-th column of the Young diagram $\mu$, numerated from left to right.  
	Then
	\begin{eqnarray}
		h^{\lambda \oplus_v \tau}_i = h^\lambda_i + h^\tau_i .
	\end{eqnarray}
	
	2. Let $\{r^\lambda_i\}$ be the (unordered) multiset of row lengths of $\lambda$.  
	Similarly, let $\{r^\tau_\alpha\}$ be the multiset of row lengths of $\tau$.  
	Then $\lambda \oplus_v \tau$ is the Young diagram whose row lengths are given by the union of these multisets: $\{r^\lambda_i, r^\tau_\alpha\}$.

	\section{The horizontal sum relation of sl and sp small multiplets}
	
	One can introduce the dual operation, the {\it horizontal componentwise sum} $\lambda \oplus_h \tau$, in a completely analogous way.
	
	It is naturally defined as follows: place two diagrams side by side, aligning their upper horizontal boundaries, and push the boxes of the right diagram to the left until they either meet boxes of the left diagram or reach the left boundary. Equivalently (and similarly to the vertical sum), in terms of the row lengths introduced above, one can define the horizontal sum by 
	\begin{eqnarray}
		r^{\lambda \oplus_h \tau}_i = r^\lambda_i + r^\tau_i .
	\end{eqnarray}
	Likewise, if $\{h^\lambda_i\}$ is the unordered multiset of column heights of the Young diagram $\lambda$, and $\{h^\tau_\alpha\}$ is the same for $\tau$, then $\lambda \oplus_h \tau$ is the Young diagram whose column heights are given by the union of these multisets: $\{h^\lambda_i, h^\tau_i\}$. 
	
	In the same way as in \cite{Mkrtchyan2025b}, one can check that statements analogous to those for the vertical sum hold for the $sl$--$sp$ relation via the horizontal sum, i.e. that for each known universal multiplet there are associated representations of $sl$ and $sp$ connected by the horizontal sum relation. As in \cite{Mkrtchyan2025b}, one may conjecture that for all other, currently unknown, multiplets, as well as the universal Casimir multiplets, there always exists such a pair of representations. One could also use the relation
	
	\begin{eqnarray}
		\lambda \oplus_h \tau=\left( \lambda^T \oplus_v\tau^T \right)^T
	\end{eqnarray}
	
	to connect the statements for horizontal sum and the vertical sum through $so(N)\sim sp(-N)$ duality.

	\section{Application of method to adjoint representation}
	
	We now illustrate the proposed approach for computing the $\gamma$-independent factors in universal quantum dimension formulae, using the adjoint representation as the simplest example. 
	
	Consider the universal formula \ref{cad} for the quantum dimension of the adjoint representation. Suppose we know only its general structure, namely that there are two $\gamma$-independent sines in the denominator, and we want to reconstruct them from the quantum dimensions of adjoints of $sl$, $so$, and $sp$. First note that this is a universal formula of the required type, i.e. it is nonzero for all three classical families. Next, write the quantum dimensions of all three: 
	
	For $sl(N)$:
	\begin{eqnarray}\label{qsl}
		\frac{\sinh\left(\frac{x}{2}(N-1)\right) \sinh\left(\frac{x}{2}(N+1)\right)}{\sinh\left(\frac{x}{2}\right)\sinh\left(\frac{x}{2}\right)}
	\end{eqnarray}
	
	For $so(N)$:
	
	\begin{eqnarray}\label{qso}
		\frac{\sinh\left(\frac{x}{4}N\right) \sinh\left(\frac{x}{2}(N-1)\right) \sinh\left(\frac{x}{2}(N-4)\right)}{\sinh\left(\frac{x}{2}\right)\sinh\left(\frac{x}{1}\right)\sinh\left(\frac{x}{4}(N-4)\right) }
	\end{eqnarray}
	
	For $sp(N)$:
	
	\begin{eqnarray}\label{qsp}
		\frac{\sinh\left(\frac{x}{8}N\right) \sinh\left(\frac{x}{4}(N+1)\right) \sinh\left(\frac{x}{4}(N+4)\right)}{\sinh\left(\frac{x}{4}\right)\sinh\left(\frac{x}{2}\right)\sinh\left(\frac{x}{8}(N+4)\right) }
	\end{eqnarray}

	Now imagine we are given these three expressions and want to recover the universal expression (\ref{cad}), or more precisely its $\gamma$-independent denominator factors. From these expressions we see that at least two sines in the denominator must be $\gamma$-independent. Assume the minimal possibility, namely two, and parameterize them as 
	
	\begin{eqnarray}\label{uqs}
		\sinh\left(\frac{x}{4}(x_1 w_x+y_1 w_y)\right)\sinh\left(\frac{x}{4}(x_2 w_x+y_2 w_y)\right)  \\
		w_x=-\frac{2\alpha+\beta}{2} \\
		w_y=\frac{\alpha+\beta}{2}
	\end{eqnarray}
	
	A minor subtlety is that we may consider all arguments of $\sinh$ to be positive (for $x>0$), since changing the sign only changes the overall sign of the entire expression (\ref{uqs}). 
	
	The values of $w_x,w_y$ for different classical algebras are given in Table \ref{tab:wxwy}. Specializing (\ref{uqs}) to $sl$ gives

	\begin{eqnarray}
		\sinh\left(\frac{x}{4}(x_1)\right)\sinh\left(\frac{x}{4}(x_2)\right) 
	\end{eqnarray}
	and comparing with (\ref{qsl}) we deduce $x_1=x_2=2$. 
	
	Next, specialize (\ref{uqs}) to $so$ and obtain 
	
	\begin{eqnarray}
		\sinh\left(\frac{x}{4}(y_1)\right)\sinh\left(\frac{x}{4}(y_2)\right) 
	\end{eqnarray}
	
	Comparing with (\ref{qso}), we deduce that either $y_1=2, y_2=4$, or vice versa. Since $x_1=x_2$, we are free to choose the labeling; we take $y_1=2, y_2=4$.
	
	In this example we found all $x_i,y_i$ from the $sl$ and $so$ data without using $sp$. This will not be the case in the more complicated example below. Here we can use the $sp$ data as a consistency check. Specialize (\ref{uqs}) to $sp$:
	
	\begin{eqnarray}
		\sinh\left(\frac{x}{4}(x_1 \frac{3}{2}-y_1 \frac{1}{2})\right)\sinh\left(\frac{x}{4}(x_2 \frac{3}{2}-y_2 \frac{1}{2})\right) 
	\end{eqnarray}
	and comparing with (\ref{qsp}), we deduce that either $x_1 \frac{3}{2}-y_1 \frac{1}{2}=1, x_2 \frac{3}{2}-y_2 \frac{1}{2}=2$, or vice versa. The second case is satisfied by our solution, so the check is successful, and our universal denominator reproduces the denominators of the quantum dimensions for all classical algebras (and exceptional ones as well, since we already know the final universal result).
	
	In general one needs the $sp$ data to determine the pairing between the multisets $\{x_i\}$ and $\{y_i\}$, since in the $sl$ case one only deduces the multiset $\{x_i\}$ and in the $so$ case only the multiset $\{y_i\}$, while the correspondence between them remains unknown. The role of the $sp$ data is to fix this pairing, since $x_i$ and $y_i$ enter the $sp$ specialization simultaneously. Moreover, it is not guaranteed that $sp$ data determines the pairing uniquely; at the very least, it imposes strong restrictions, leaving only a small number of possibilities.

	\section{Application of method: E representation}
	
	Consider the universal dimension formula presented in \cite{AvetisyanIsaevKrivonosMkrtchyan2024} for the universal multiplet E: 
	
	\small
	\begin{eqnarray}\label{dimE}
		dim E=-\frac{64 (\alpha+\gamma) (2 \alpha+\gamma) (\alpha+2 \gamma) (\beta+\gamma) (2 \beta+\gamma) (\beta+2 \gamma) (\alpha+\beta+\gamma)  }{\alpha^2 \beta^2 \gamma (\alpha-3 \beta) (\alpha-\beta)^2 (3 \alpha-\beta) (\alpha-\gamma) (\beta-\gamma) (\alpha+\beta-2 \gamma)} \times \\
		(2 \alpha+\beta+\gamma)(\alpha+2 \beta+\gamma) (2 \alpha+2 \beta+\gamma) (2 \alpha+\beta+2 \gamma) (\alpha+2 \beta+2 \gamma)
	\end{eqnarray}
	\normalsize
	
	This expression is symmetric under the swap $\alpha \leftrightarrow \beta$, so the corresponding small universal multiplets contain three representations, although most of them are zero, as we shall see. One member of the $sl$ small multiplet is
	
	\begin{eqnarray}
		D_s(\ydiagram{3,1},\ydiagram{2,1,1})=(2,1,0,...0,1,0,1)_s
	\end{eqnarray}
	
	with dimension
	
	\begin{eqnarray}
		\frac{1}{32} (N-4) (N-2) (N-1)^2 (N+1)^2 (N+2) (N+4)
	\end{eqnarray}
	
	It is invariant under $N \leftrightarrow -N$ due to the symmetry of (\ref{dimE}) under $\alpha \leftrightarrow \beta$.

	The other two members of the $sl$ small multiplet are zero. 
	
	The corresponding $so$ associated representation has dimension and diagram 
	\begin{eqnarray}
		&\frac{1}{630} (N-6) (N-4) (N-2) (N-1) N^2 (N+2) (N+4) \\
		&\ydiagram{3,2,1,1,1}
	\end{eqnarray}
	with the other elements of the $so$ small multiplet being zero. 
	
	The corresponding data for the $sp$ multiplet are 
	
	\begin{eqnarray}
		&\frac{1}{630} (N-4) (N-2) N^2 (N+1) (N+2) (N+4) (N+6) \\
		&\ydiagram{5,2,1}
	\end{eqnarray}
	
	and other members are zero. 
	
	Next we consider the quantum dimensions of these representations. 
	The quantum dimension of the $sl$ representation is obtained by $q$-deforming all factors in the hook formula:
	
	\tiny
	\begin{eqnarray}\label{qsle}
		\dim_q(\lambda)=
		\frac{
			\sinh\!\big(\frac{x}{2}(N-4)\big)\,\sinh\!\big(\frac{x}{2}(N-2)\big)\,\sinh\!\big(\frac{x}{2}(N-1)\big)^{2}\,
			\sinh\!\big(\frac{x}{2}(N+1)\big)^{2}\,
		}{
			\sinh(\frac{x}{2})^{4}\,\sinh(x)^{2}\,\sinh(2x)^{2} 
		} \times \\  \nonumber
		\sinh\!\big(\frac{x}{2}(N+2)\big)\,\sinh\!\big(\frac{x}{2}(N+4)\big)
	\end{eqnarray}
	
	\normalsize
	
	The quantum dimension of the $so(2n)$ representation is 
	
	\tiny
		
		\begin{eqnarray}\label{qsoe}
		\dim_q=
		\frac{
			\sinh(\frac{x}{2}n)\,\sinh(\frac{x}{2}(n+2))\,
			\sinh(x(n-5))\,\sinh(x(n-3))\,\sinh(x(n-2))\,
		}{
			\sinh(\frac{x}{2})^{3}\,\sinh(x)\,\sinh(3\frac{x}{2})^{2}\,\sinh(5\frac{x}{2})\,\sinh(7\frac{x}{2})\,\sinh(\frac{x}{2}(n-5))\,\sinh(\frac{x}{2}(n-1))
		} \times \\  \nonumber
		\sinh(x(n-1))^{2}\,
		\sinh(\frac{x}{2}(2n-1))\,\sinh(x n)\,\sinh(x(n+1))
	\end{eqnarray}
	
	\normalsize
	
	Similarly, for the $sp(2n)$ representation:
	
	\tiny
	
		\begin{eqnarray}\label{qspe}
		\dim_q=
		\frac{
			\sinh(\frac{x}{4}(n-2))\,\sinh(\frac{x}{4}n)\,\sinh(\frac{x}{2}(n-1))\,\sinh(\frac{x}{2}(n))\,\sinh(\frac{x}{4}(2n+1))\,
		}{
			\sinh(\frac{x}{4})^{3}\,\sinh(\frac{x}{2})\,\sinh(3\frac{x}{4})^{2}\,\sinh(5\frac{x}{4})\,\sinh(7\frac{x}{4})\,\sinh(\frac{x}{4}(n+1))\,\sinh(\frac{x}{4}(n+5))
		}  \times\\ \nonumber
		\sinh(\frac{x}{2}(n+1))^{2}\,\sinh(\frac{x}{2}(n+2))\,\sinh(\frac{x}{2}(n+3))\,\sinh(\frac{x}{2}(n+5))
	\end{eqnarray}
	
	\normalsize
	
	With this data we would like to determine, under natural assumptions, the $\gamma$-independent factors in the universal quantum dimension of the universal representation $E$. 
	We assume the following form of the quantum dimension of this representation. First, recall that when specialized to $sl$, the universal formula corresponds to the sum of a representation and its image under the $\mathbb{Z}_2$ automorphism of the Dynkin diagram of $sl$, with equal dimensions and quantum dimensions. Hence a factor $2$ appears in front of the product of sines. Such a factor appears in other universal quantum dimensions via the universal factor

	\begin{eqnarray}\label{z2corr}
		\frac{\sinh(x(\alpha+\beta)c/2)}{\sinh(x(\alpha+\beta)c/4)}
	\end{eqnarray} 
	where $c$ is a constant to be determined. 
	In the case of $sl$ algebras, where $\alpha+\beta=0$, this factor gives $2$; for permutations and for other algebras it gives a nonzero ratio of sines. 
	
	Next, we separate the $\gamma$-independent sine factors and assume the minimal number of $\gamma$-independent factors in the denominator, which in this case must be $8$ in order to reproduce the denominators of the above formulae. Finally, we allow a number, say $k$, of $\gamma$-dependent sines in the denominator, and correspondingly $8+k$ $\gamma$-dependent sines in the numerator. 
	
	Let us denote the $\gamma$-independent sines in the denominator by:

	\begin{eqnarray}\label{udenomE}
		\prod_{i=1}^{8} \sinh\left(\frac{x}{2}(x_iw_x+y_iw_y)\right) 
	\end{eqnarray}
	
Recall table \ref{tab:wxwy} for  values of $w_x,w_y$ for different classical algebras. 
	
	\begin{table}
		\caption{Values of $w_x, w_y$ }
		\label{tab:wxwy}
		$\begin{array}{|c|c|c|c|}
			\hline
			& sl  & so & sp  \\
			\hline
			w_x	&  1&0  & \frac{3}{2} \\
			\hline
			w_y	&0 & 1  &-\frac{1}{2}  \\
			\hline
		\end{array}$
	\end{table}

	The product (\ref{udenomE}), taken in denominator, together with  (\ref{z2corr}) should reproduce the $\gamma$-independent denominators of the quantum dimensions for $sl$, $so$, and $sp$.  
	
	For the $sl$ algebra the set of arguments of sines in the denominator of (\ref{qsle}) (we call argument of sin the numbers $u$ such that $\sinh\left(\frac{x}{2}u\right)$ occurs) is the unordered multiset
	\begin{eqnarray}
		1,1,1,1,2,2,4,4
	\end{eqnarray}
	From the other side, the set of arguments coming from the universal expression is $\{x_i, i=1,\dots,8\}$. 
	This means that one has, after a suitable reindexing,   
	
	\begin{eqnarray}
		x_1=1,x_2=1,x_3=1, x_4=1 \\
		x_5=2,x_6=2,x_7=4, x_8=4
	\end{eqnarray}
	
	Similarly, we compare the universal expression (\ref{udenomE}), specialized to $so$, with the denominator of the $so$ quantum dimension. These are
	\begin{eqnarray}\label{qdsodenom}
		1,1,1,2,3,3,5,7
	\end{eqnarray}
	
	The set of arguments from the universal expression (\ref{udenomE}) is given by $\{y_i, i=1,\dots,8\}$, with one correction. The $\gamma$-independent sine in the numerator of the factor (\ref{z2corr}) should cancel one of the $y$'s, say $y_k$ (so $(\alpha+\beta)c=2c=y_k$), and the sine in the denominator of (\ref{z2corr}) then contributes an additional factor. Since the argument of the sine in the numerator of (\ref{z2corr}) is twice the argument of the sine in the denominator, in the universal quantum dimension the multiset $\{y_i, i=1,\dots,8\}$ is effectively replaced by $\{y_1,\dots,y_{k-1}, \frac{y_k}{2},y_{k+1},\dots,y_8\}$. Note that this multiset should coincide with unordered (\ref{qdsodenom}), so one does not know the value of $y_i$ for a given $i$. To resolve this ambiguity we use additional information from the $sp$ quantum dimension. 
	
	The point is that when specialized to $sp$, the arguments in the universal quantum dimension (\ref{udenomE}) involve both $x_i$ and $y_i$ simultaneously (see Table \ref{tab:wxwy}), so one can hope to determine a unique connection between (\ref{qdsodenom}) and $\{y_i\}$. 
	
	Thus, the arguments of sines in the denominator of the quantum dimension for $sp$ are 
	
	\begin{eqnarray}
		\frac{1}{2},	\frac{1}{2},	\frac{1}{2},1,	\frac{3}{2},\frac{3}{2},\frac{5}{2},\frac{7}{2}
	\end{eqnarray}
	
	This is the same as for $so$, up to common factor $\frac{1}{2}$, since the $so$ and $sp$ diagrams are transposes of each other, so their (quantum) dimensions agree up to $n \leftrightarrow -n$ and rescaling of $x$; in particular, the $n$-independent sine factors are connected by rescaling, only. 
	
	The universal quantum dimension formula (\ref{udenomE}) gives for denominators the arguments $\{\frac{3}{2}x_i-\frac{1}{2}y_i, i=1,\dots,8\}$. However, here again one has a nontrivial contribution from (\ref{z2corr}). Since we already assume that in the $so$ case this factor equals
	
	\begin{eqnarray}\label{z2corrso}
		\frac{\sinh(x(\alpha+\beta)c/2)}{\sinh(x(\alpha+\beta)c/4)}=	\frac{\sinh(xy_k/2)}{\sinh(x y_k/4)}
	\end{eqnarray} 
	
	then for $sp$ it becomes

	\begin{eqnarray}\label{z2corrsp}
		\frac{\sinh(x(\alpha+\beta)c/2)}{\sinh(x(\alpha+\beta)c/4)}=	\frac{\sinh(xy_k/4)}{\sinh(x y_k/8)}
	\end{eqnarray} 
	
	Accordingly, one of the numbers in $\{\frac{3}{2}x_i-\frac{1}{2}y_i\}$ should equal $y_k/2$, and it is replaced by $y_k/4$ as an argument of sines. 
	
	After an exhaustive case analysis, we find that $y_k=2$, and then the unordered multiset of $y_i$ becomes 
	
	\begin{eqnarray}
		1,1,2,2,3,3,5,7
	\end{eqnarray}
	
	The unordered multiset $\{\frac{3}{2}x_i-\frac{1}{2}y_i, i=1,\dots,8\}$ becomes 
	
	\begin{eqnarray}
		\frac{1}{2},	\frac{1}{2},1,1, \frac{3}{2},\frac{3}{2},\frac{5}{2},\frac{7}{2}
	\end{eqnarray}
	
	Thus one should order the list of $y_i$ so that the corresponding list $\{\frac{3}{2}x_i-\frac{1}{2}y_i, i=1,\dots,8\}$ matches the multiset above. 
	One finds a unique solution:
	
	\begin{eqnarray}
		x_1=1, y_1= 1,   \frac{3}{2}x_1-\frac{1}{2}y_1=1\\
		x_1=1, y_2= 1,   \frac{3}{2}x_2-\frac{1}{2}y_2=1\\	
		x_3=1,y_3=2,  \frac{3}{2}x_3-\frac{1}{2}y_3=\frac{1}{2} \\	
		x_4=1,y_4=2,  \frac{3}{2}x_4-\frac{1}{2}y_4=\frac{1}{2} \\	
		x_5=2,y_5=3,  \frac{3}{2}x_5-\frac{1}{2}y_5=\frac{3}{2} \\
		x_6=2,y_6=3,  \frac{3}{2}x_6-\frac{1}{2}y_6=\frac{3}{2} \\
		x_7=4,y_7=5,  \frac{3}{2}x_7-\frac{1}{2}y_7=\frac{7}{2} \\
		x_8=4, y_8=7, \frac{3}{2}x_8-\frac{1}{2}y_8=\frac{5}{2} 
	\end{eqnarray}
	
	Finally, with this solution for $x_i,y_i$ we compute the universal denominator (\ref{udenomE}): 
	
	\begin{eqnarray}\label{udenomE2}
		\prod_{i=1}^{8} \sinh\left(\frac{x}{2}(x_iw_x+y_iw_y)\right) =\\
		\sinh\left(\frac{x}{4}\alpha\right)^2 	\sinh\left(\frac{x}{4}\beta\right)^2 	\sinh\left(\frac{x}{4}(\beta-\alpha)\right)^2 \times \\	\sinh\left(\frac{x}{4}(\beta-3\alpha)\right)\sinh\left(\frac{x}{4}(3\beta-\alpha)\right)
	\end{eqnarray}
	
	In the limit $x \rightarrow 0$, i.e. in the universal dimension formula, this is proportional to 
	
	\begin{eqnarray}
		\alpha^2 \beta^2(\beta-\alpha)^2(\beta-3\alpha)(3\beta-\alpha)
	\end{eqnarray}
	
	which exactly coincides with the universal dimension formula (\ref{dimE}). Note that our derivation is completely different from that in \cite{AvetisyanIsaevKrivonosMkrtchyan2024}, so this agreement is a nontrivial confirmation of the present approach.   
	
	\section{Conclusion}
	
	We proposed a method for deriving universal quantum dimension formulae from standard quantum dimension data for classical simple Lie algebras. At present, it is restricted to universal formulae whose associated $sl$, $so$, and $sp$ representations are all nonzero. There are series of such universal formulae (e.g. Cartan powers of the adjoint), as well as an opposite examples (Cartan powers of $X_2$). The method also has an ambiguity: it is not clear a priori whether the $sp$ data determines a unique correspondence between the multisets of coefficients $\{x_i\}$ and $\{y_i\}$. At the very least, it seems to leave only a small number of possibilities. 
	
	We tested this method for the universal adjoint multiplet and derived a (new) formula for the $\gamma$-independent factors of the universal quantum dimension of the $E$ multiplet. 
	
	An advantage is the small number of calculations, which makes large representations affordable. Hopefully, this method may lead to a deeper understanding of the structure of universal formulae. 
	
	To make this program fully effective, one must extend this method on computation of the $\gamma$-dependent factors of universal quantum dimension formulae. This will be a little bit more difficult, since the pattern of  cancellation of factors between numerator and denominator can be complicated; this work is in progress. Perhaps in this way one can obtain all possible universal quantum dimension formulae with non-zero associated $sl, so, sp$ representations. Also out of scope of this approach are the quantum dimension formulae for Casimir subspaces consisting from a several irreps, since their universal dimensions are not the product/ratio of linear factors, but ratio of some more complicated polynomials.
	
	\section{Acknowledgments}
	
	This work was partially supported by the Science Committee of the Ministry of Science and Education of the Republic of Armenia under contracts 21AG-1C060 and 24WS-1C031.

	\appendix
	
	\section{Vogel's table}
	
	Below we reproduce Vogel's table of points in the projective plane corresponding to simple Lie algebras. The homogeneous coordinates $\alpha, \beta, \gamma$ on the projective plane are relevant up to a common rescaling and permutations.
	
	\begin{table}[h] \caption{Vogel's parameters for simple Lie algebras}     \label{tab:Vogel}
		\begin{tabular}{|r|r|r|r|r|r|} 
			\hline Algebra/Parameters & $\alpha$ &$\beta$  &$\gamma$  & $t=\alpha+\beta+\gamma$ & Line \\ 
			\hline $sl(N)$ & -2 & 2 & $N$ & $N$ & $\alpha+\beta=0 $\\ 
			\hline $so(N), sp(-N)$ & -2  & 4 & $N-4$ & $N-2$ & $\alpha+2\beta=0$ \\ 
			\hline $Exc(n)$ & -2 & $n+4$  & $2n+4$ & $3n+6$& $\gamma=2(\alpha+\beta)$ \\ 
			\hline 
		\end{tabular} 
	\end{table}
	
	In Table \ref{tab:Vogel}, for $sl(N)$ and $so(N)$, $N$ is a positive integer; for $sp(-N)$, $N$ is a negative even integer; for the exceptional line $Exc(n)$, one has $n=-1,-2/3,0,1,2,4,8$ corresponding to $A_2,G_2, D_4, F_4, E_6, E_7, E_8$, respectively.

\end{document}